\documentclass[pra,twocolumn,showpacs,amsmath,amssymb]{revtex4}
\usepackage{graphicx}
\usepackage{bm}

\begin{document}
\title{Solving the $\bm m$-mixing problem for the three-dimensional time-dependent
Schr\"{o}dinger equation by rotations: application to strong-field
ionization of H$_2{}^+$}

\author{T. K. Kjeldsen}

\author{L. A. A. Nikolopoulos}

\author{L. B. Madsen}

\affiliation{Lundbeck Foundation Theoretical Center for Quantum System Research,
Department of Physics and Astronomy, University of Aarhus, 8000 Aarhus
C, Denmark.}


\begin{abstract}
  We present a very efficient technique for  solving the three-dimensional
  time-dependent Schr\"{o}dinger equation. Our method is applicable  to a wide range of
  problems where a fullly three-dimensional solution is required, i.e., to cases 
  where no symmetries exist that reduce the dimensionally of the problem. Examples include 
  arbitrarily oriented molecules in external fields and atoms interacting with
  elliptically polarized light. We demonstrate that even in such cases, the 
  three-dimensional problem can be decomposed exactly into two two-dimensional problems
  at the cost of introducing  a trivial rotation transformation. We supplement the theoretical
  framework with numerical results on strong-field ionization of arbitrarily oriented
  H$_2{}^+$ molecules.
\end{abstract}
\maketitle

\section{Introduction}
In atomic physics the spherical symmetry of atoms promotes
the spherical coordinates to a special position. The three independent
variables are $(r,\theta,\phi)$, with $r$ the radial distance of the
electron with respect to the nucleus, $\theta$ the polar angle and
$\phi$ the azimuthal angle. The Schr\"{o}dinger equation for the
hydrogen atom is separable in these coordinates with wave functions of the form
$\psi_{nlm}(\bm r) = R_{nl}(r)Y_{lm}(\theta,\phi)$ separated into
a radial wave function $R_{nl}(r)$ and a spherical harmonic
$Y_{lm}(\theta,\phi)$. Furthermore, configurations of this type with
associated orbitals $\psi_{nlm}(\bm {r})$ form the building blocks of
Slater determinants and consequently of mean field approaches to atomic
structure.
Even for molecules where the presence of
multiple nuclei breaks the spherical symmetry ($\left[L^2,H\right] \neq
0$), single-centre expansions in spherical harmonic basis has been used
successfully~\cite{martin:1999}.  

For a general problem involving a single active electron we are thus led to
the consideration of the three-dimensional  time-dependent Schr\"{o}dinger
equation in spherical coordinates and for the reduced wave function ($\Phi =
r \Psi$) we seek a solution of the form 
\begin{equation}
  \Phi(\bm {r},t)=
  \sum_{l = 0}^\infty \sum_{m=-l}^lf_{lm}(r,t)Y_{lm}(\theta,\phi).
\label{eq:wf_lm}
\end{equation}
A very important advantage of this representation is that we can benefit
from angular momentum theory when dealing with the angular degrees of
freedom.  An outstanding problem, however, remains. The problem, which is
referred to as the $m$-mixing problem among computational scientists, is
that often couplings---external or internal---are present that introduce a
mixing of $m$'s across $l$'s. Such $m$-mixings occur for example when an
atom is subject to an elliptically polarized field or to a linearly
polarized field described beyond the dipole approximation. When $m$ is no
longer  conserved, the dynamics affects all three coordinates and  a
numerical simulation is difficult: three-dimensional calculations tend to be
extremely time-consuming and computationally demanding.

In the course of our recent work concerned with alignment-dependent response
of molecules to strong external fields we found a solution that speeds up
the calculation by the use of an exact mapping of the three-dimensional
problem to two two-dimensional problems. In the following we discuss the
method by the specific example of the response of an arbitrarily oriented
diatomic molecule to an external perturbation so strong that the system is
ionized. As will become clear, the central ideas are completely general and
carry over to the related case of atoms in elliptically polarized fields,
polyatomic molecules as well as $m$-problems in geology and astronomy  where
expansions in spherical harmonics are also often encountered.

The paper is organized as follows: in Sec.~\ref{sec:general} we give an
overview over the basic idea of our technique. In Sec.~\ref{sec:calculation}
we outline the numerical implementation and discuss physical results for
H$_2{}^+$ strong-field ionization.  Sec.~\ref{sec:conclusion} concludes.

\section{Basic ideas and principles}
\label{sec:general}

\begin{figure}
  \begin{center}
    \includegraphics[width=0.9\columnwidth]{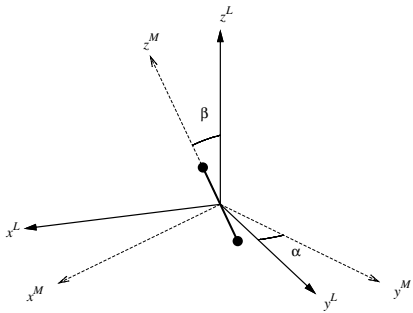}
  \end{center}
  \caption{The orientation of the molecular coordinate
  system ($M$, dashed) with respect to the laboratory fixed system
  ($L$, solid). In the figure only the Euler angles $\alpha,\beta$ are nonvanishing.}
  \label{fig:euler}
\end{figure}
We illustrate the basic ideas and principles of the method by discussing the
specific example of a linear diatomic molecules in an external
electromagnetic field.  In Fig.~\ref{fig:euler} we show the coordinate
systems which are relevant for the field-molecule problem. The coordinates
$(x^L,y^L,z^L)$ specify the laboratory ($L$) fixed coordinate system defined
by the external polarization vector. We assume that the field is linearly
polarized and return to the generalization to elliptically polarized light
in Sec.~\ref{sec:conclusion}. The coordinate system denoted by superscripts
$M$ is the molecular fixed frame and is rotated by the Euler angles
$(\alpha,\beta,\gamma)$ with respect to the laboratory fixed system. The
rotation is accomplished by an $\alpha$ rotation around the $z^L$-axis,
followed by a $\beta$ rotation around the $y^M$-axis, and finally a $\gamma$
rotation around the $z^M$-axis.  For the case considered the only really
distinct geometries are associated with the angle $\beta$.  Results for
different orientations due to the angle $\alpha$ are trivially related by a
simple rotation around the $z^L$ axis. Also the $\gamma$ rotations around
the molecular axis are insignificant as a consequence of the axial symmetry
of the molecule.

We want to determine how the wave function of an electron is affected by the
operators $V^{(M)}$ and $V^{(I)}(t)$, corresponding to the interaction with
the nuclei and the field, respectively.  We assume that we can treat these
two operators separately, which is the case in a split-operator approach as
described in Sec.~\ref{sec:calculation} below.  Our strategy is first to
represent the wave function in the molecular frame and calculate the action
of $V^{(M)}$. Secondly, we transform the updated wave function to the
laboratory fixed frame and apply the operator $V^{(I)}$. Finally we can
return to the molecular frame by the inverse rotation.
\begin{figure}
\includegraphics[width=\columnwidth]{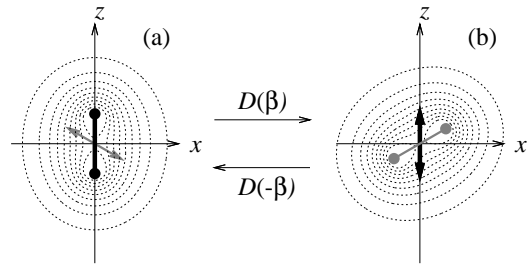}
\caption{
Schematic picture of the rotation operation. The contour lines indicate the
field free $1\sigma_g$ ground state of H$_2{}^+$ in the $xz$ plane.  The
double headed arrow shows the direction of the laser polarization vector. In
(a) we calculate the action of the molecular potential and express the wave
function in the molecular frame with the internuclear axis parallel to the
$z$ axis. In panel (b) we transform the wave function to the laboratory
fixed system with the laser polarization parallel to the $z$ axis in order
to propagate by the field interaction. The transformation between the two
frames is represented by the rotation operator $D$.
}
\label{fig:rotation}
\end{figure}
These forward ($\beta$) and backward ($-\beta$) rotations of the wave
function are illustrated in Fig.~\ref{fig:rotation}. The active interaction
($V^{(M)}$ or $V^{(I)}$) is marked by black and the inactive operation is
gray.  This propagation scheme for arbitrary orientation of the polarization
axis with respect to the internuclear axis, exhibits the strength of the
present approach since it allows us to perform the calculations very
efficiently.  Whenever we apply an axially symmetric operator, we do not mix
different $m$ states provided that the wave function is expressed in the
proper reference frame. Thus we can apply the operator separately on each
different $m$ state. The decoupling of different $m$ states means
effectively that we have reduced the three-dimensional problem to a number
of two-dimensional problems in addition to two rotation operations.

The rotation transformation is in principle possible in all sets of
coordinates and the separation in $m$ applies to any coordinate system where
the azimuthal angle $\phi$ is an independent variable, e.g. cylindrical,
parabolic, or spheroidal coordinates.  The two unique features of the
spherical representation \eqref{eq:wf_lm} are that (i) the transformation
matrix contains Wigner rotation functions which are known analytically and
(ii) the transformation is guaranteed to be exactly unitary for functions
that are bandwidth limited by a maximum $l=l_\text{max}$, i.e., the
population in states with $l > l_\text{max}$ is zero.

\section{Numerical results}
\label{sec:calculation}
In the present work, we solve the time-dependent Schr\"{o}dinger equation
(TDSE) for the electronic motion in H$_2{}^+$ in the presence of a
time-dependent electromagnetic field. We represent the angular variables in
a basis of spherical harmonics and write the reduced wave function as in
Eq.~\eqref{eq:wf_lm}.  The radial functions $f_{lm}$ which contain the time
dependence are discretized on an equidistant spatial mesh. The expansion in
spherical harmonics is truncated such that $l \le l_\text{max}$ leading to a
total number of $(l_\text{max}+1)^2$ angular basis functions.  The reduced
wave function satisfies the TDSE with the Hamiltonian [atomic units
$(\hbar=|e|=m_e=a_0 =1)$ are used throughout]
\begin{equation}
  H(t) = -\frac{1}{2}\frac{\partial^{2}}{\partial r^{2}} +
  \frac{L^{2}}{2r^{2}} +
  V(t)  = T_r + T_l + V(t),
  \label{eq:hamilton}
\end{equation}
where $L$ is the usual angular momentum operator and $V$ includes
the electronic interaction with the field and the nuclei.
We solve the time-evolution from time $t$ to $t+\tau$
numerically by using the split-operator technique 
\begin{equation}
\Phi({\bm
r},t+\tau)=
e^{-iT_{r}\frac{\tau}{2}}e^{-iT_{l}\frac{\tau}{2}}e^{-iV(t+\frac{\tau}{2})\tau}e^{-iT_{l}\frac{\tau}{2}}e^{-iT_{r}\frac{\tau}{2}}
\Phi({\bm r},t).
  \label{eq:splitop}
\end{equation}
The error in the propagation scheme above is approximately cubic in $\tau$
and occurs mainly due to the splitting of non-commuting operators.  A
related propagation scheme was applied in geometries with azimuthal symmetry
\cite{hermann:1988}, and the propagation techniques used for the kinetic
operators  $T_r$ and $T_l$ are readily extended to our fully
three-dimensional problem.  We will therefore turn to the new propagation
method of the molecular potential and the field interaction.

We describe the electromagnetic field in the dipole approximation 
by the vector potential
\begin{equation}
  \bm{A}(t) = {\bm {\hat{e}}} A_0(t) \cos(\omega t),
\label{eq:At}
\end{equation}
where $A_0(t)$ is the envelope function, $\omega$ the frequency and
${\bm {\hat{e}}}$ the polarization direction. The electric field is 
obtained as $\bm {F}(t) = -d \bm {A}(t)/dt$.
The operator $V$ in Eq.~\eqref{eq:hamilton} is written as the sum of the
field interaction and the molecular potential
\begin{equation}
  V(t) = V^{(I)}_{r,\theta_L}(t) + V^{(M)}_{r,\theta_M},
\end{equation}
where the subscripts denote the variables on which the operators act.
$\theta_M$ is the polar angle in the molecular frame
[Fig.~\ref{fig:rotation}~(a)] and $\theta_L$ the polar angle in the laboratory
fixed system [Fig.~\ref{fig:rotation}~(b)]. The molecular operator is diagonal
in coordinate space
\begin{equation}
  V^{(M)}_{r,\theta_L} = V^{(M)}(r,\theta_L),
\end{equation}
while the field interaction can be represented either in the length- (LG) or the
velocity gauge (VG) as
\begin{widetext}
\begin{equation}
  V^{(I)}_{r,\theta_L}(t) = \left\{ 
  \begin{array}{lr}
    F(t) r \cos \theta_L & \text{LG} \\
   iA(t)\left[  \frac{1}{r}\left( \cos \theta_L + \sin \theta_L
  \frac{\partial}{\partial \theta_L}\right) - \cos \theta_L
  \frac{\partial}{\partial r} \right] & \text{VG}
  \end{array}
  \right..
\end{equation}
\end{widetext}
To calculate the action of $V$ in the propagation we make the split 
\begin{equation}
  e^{-iV(t+\frac{\tau}{2})\tau} \approx e^{-iV^{(M)}\tau/2}
  e^{-iV^{(I)}(t+\frac{\tau}{2})\tau} e^{-iV^{(M)}\tau/2}.
\end{equation}
For each radial grid point $r_i$ we write the wave function as a vector
in the spherical harmonics basis, cf. Eq.~\eqref{eq:wf_lm}
\begin{equation}
 \bm {f}^{(M)}(r_i,t) = 
 \begin{pmatrix}
   f^{(M)}_{00}(r_i,t)\\
    f^{(M)}_{10}(r_i,t)\\
    \vdots\\
    f^{(M)}_{l_\text{max},0}(r_i,t)\\
   \hline
    f^{(M)}_{11}(r_i,t)\\
    \vdots\\
    f^{(M)}_{l_\text{max},1}(r_i,t)\\
    \hline
    \vdots\\
 \end{pmatrix},
 \label{eq:f_lm_matrix}
\end{equation}
where the coefficients refer to the molecular frame.  The molecular
potential is diagonal in the radial coordinate, and cannot induce mixings
vectors that belong to different radial coordinates.  We evaluate the action
of $e^{-iV^{(M)}\tau/2}$ by its matrix representation in the spherical
harmonics basis for each fixed value of $r$
\begin{equation}
  \langle lm | e^{-iV^{(M)}\tau/2} | l'm'\rangle =
  \delta_{mm'} \langle lm | e^{-iV^{(M)}(r,\theta_M)\tau/2 }| l'm\rangle.
\end{equation}
The selection rule $m=m'$ occurs since $V^{(M)}$ is independent of $\phi_M$.
Now it is evident that $e^{-iV^{(M)}\tau/2}$ is represented by
a the block diagonal form
\begin{equation}
  \begin{pmatrix}
    \left[
    \begin{smallmatrix}
      m = 0\\
      l = 0,1,2,\cdots,l_\text{max}
      \vspace{0.5cm}
    \end{smallmatrix}\right] & &  & \text{\Large 0} \\
    
    &  \left[
    \begin{smallmatrix}
      m = 1\\
      l = 1,2,\cdots,l_\text{max}
      \vspace{0.3cm}
    \end{smallmatrix}\right] & & \\
          & & \ddots & \\
    \text{\Large 0}& & & 
    \left[
    \begin{smallmatrix}
      m = l_\text{max}\\
      l = l_\text{max}
    \end{smallmatrix}\right]
  \end{pmatrix}.
\end{equation}
Although not
essential for our present discussion, we note that for inversion
symmetric potentials as in the case of H$_2{}^+$, a further block
diagonalization in even and odd parity blocks can be obtained.  
From the block diagonal structure of the matrix representation, it is clear
that the propagation can be accomplished separately within each $m$ subspace, 
and the full three-dimensional propagation effectively reduces to independent 
two-dimensional propagations, which can be solved by matrix multiplications
on each $m$ block. 
There is a total number of $2l_\text{max}+1$ individual $m$ blocks with 
dimensionality between $1$ and $l_\text{max}+1$.

After having applied the molecular potential we 
transform the wave function to the laboratory fixed frame. We relate the
expansion in spherical harmonics in different frames by  representation of
the rotation operator in spherical harmonics, i.e., the Wigner
rotation matrix $\bm {D}(\alpha,\beta,\gamma)$. The laboratory fixed expansion
coefficients are then obtained as $\bm {f}^{(L)}(r_i,t) =
\bm {D}(\alpha,\beta,\gamma)\cdot \bm {f}^{(M)}(r_i,t)$. 
We note that this matrix multiplication is very fast since the rotation 
does not mix different $l$'s and $\bm {D}(\alpha,\beta,\gamma)$ is
consequently sparse.
Also note that the rotation operation is independent of the radial coordinate and
we can therefore use the same rotation operation on all the
vectors~\eqref{eq:f_lm_matrix} for different $r$'s.

Having obtained the wave function in the laboratory fixed frame, we can
easily apply the field interaction operator.  Again, without $m$-couplings,
the individual two-dimensional problems can be solved
straightforwardly~\cite{bauer:2006}.  Finally we return to the molecular
frame by the inverse transformation $\bm {f}^{(M)}(r_i,t) = \bm
{D}^\dagger(\alpha,\beta,\gamma)\cdot \bm {f}^{(L)}(r_i,t)$. 

We close this section with a few remarks on the scaling of the computations
with the size of the problem. In an alternative three-dimensional approach
where we in a single step treat the total $V$ and mix between all
$(l_\text{max}+1)^2$ angular basis states, the computational complexity
scales as $O(l_\text{max}^4)$~\cite{hansen:031401}. Our present method, on
the other hand, scales more favourably as $O(l_\text{max}^{2.7})$.  In
numerical simulations  for typical bandwidths of $l_\text{max}\sim 15-39$,
we have checked that both three-dimensional methods agree in their
predictions but with a great speed-up of the order of a factor of $100-500$
in favor of the new method.

\subsection{Ionization of H$_2{}^+$}
We calculate the ionization probability for H$_2{}^+$ induced by a strong
infra-red light source.  The two protons are fixed at the equilibrium
internuclear distance of $2\, \mbox{a.u.}$.  The field is taken to be
linearly polarized with frequency $\omega = 0.057\, \mbox{a.u.}\, (\lambda =
800\, \mbox{nm})$, and peak intensity $5\times 10^{14}\, \mbox{W/cm}^2$.  We
use a sine-square envelope function that encloses seven optical cycles,
corresponding to a total pulse duration of $19\, \mbox{fs}$.  Convergent
results are obtained  with $l_\text{max} = 23$ and $1024$ radial grid points
extending to a box size of $150\, \mbox{a.u.}$. In order to avoid
reflections at the edge of the box, we impose an absorbing boundary.  The
time step size is $\tau = 5\times10^{-3}\, \mbox{a.u.}$.  We choose the
velocity gauge form of the interaction since it is superior to the length
gauge in producing converged results for dynamical
problems~\cite{cormier:1996a,kamta:2005}.

First we calculate the angular differential ionization probability. For that
purpose we need the gauge invariant current density
\begin{equation}
  \bm J(\bm r,t) = \textrm{Re}\left[ \Psi^*(\bm r,t) \left(\bm p + \bm
  A(t) \right) \Psi(\bm   r,t)\right],
  \label{eq:flux}
\end{equation}
where $\bm p = -i\nabla$ is the canonical momentum.
We relate the outgoing radial probability flux at some large distance $\mathcal{R}$ to
the differential ionization probability in the laboratory fixed frame
\begin{equation}
  \frac{dP}{d\Omega_L} = \int_0^\infty dt \,\bm{\hat r}_L \cdot \bm
  J(\mathcal{R},\Omega_L,t) 
  \mathcal{R}^2.
  \label{eq:diff_ionization}
\end{equation}
We must of course choose $\mathcal{R}$ to be smaller than the radial distance at which
we turn on the absorbing potential.
\begin{figure}
  \begin{center}
    \includegraphics[width=0.7\columnwidth]{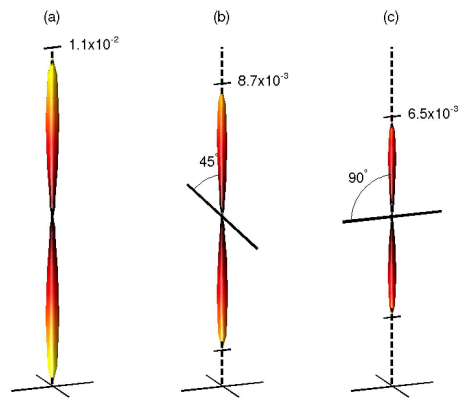}
  \end{center}
  \caption{(Color online) Angular differential ionization probability
  $dP/d\Omega_L$ for the 
  alignment angles (a) $0^\circ$, (b) $45^\circ$, and (c) $90^\circ$. 
  The laser polarization direction is vertical in all panels and the molecular axis is
  indicated by the thick solid line. 
  The numbers on the axes indicate $dP/d\Omega_L|_{\theta_L = 0}$. 
  The parameters of the electromagnetic
  field are specified in the text.}
  \label{fig:ion_angular}
\end{figure}
Figure~\ref{fig:ion_angular} shows the angular differential probabilities
for the alignment angles $0^\circ$, $45^\circ$, and $90^\circ$. In all
cases, the electron escapes exclusively in a very narrow cone along the
polarization direction.  These results are in accordance with expectations
from the quasistatic tunneling picture.  The ionization dynamics is often
considered as being tunneling-like for strong, low frequency fields where
the Keldysh parameter fullfils $\gamma < 1$~\cite{Keldysh}. In the present
case $\gamma = 0.7$ at the peak intensity.  In the tunneling picture the
electron is assumed to escape near the field direction since the barrier has
its shortest spatial extension in that direction~\cite{smirnov66}.

\begin{figure}
  \begin{center}
    \includegraphics[width=0.8\columnwidth]{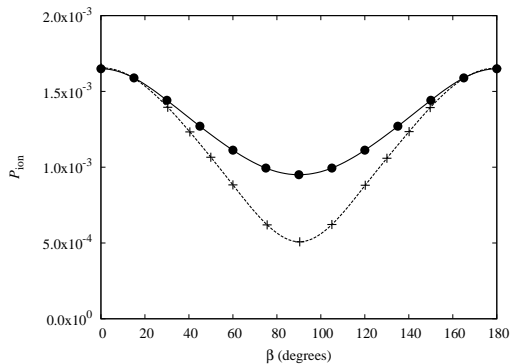}
  \end{center}
  \caption{Total ionization probability as a function of alignment angle.
  The present results are indicated by the solid line. The dashed line is
  taken from Ref.~\cite{kamta:2005} after scaling by the factor $0.18$.}
  \label{fig:ion_tot}
\end{figure}

The most notable difference between panels (a)-(c) is the overall scaling of
the distribution which decreases with increasing angle between the
polarization and internuclear axes. We can qualitatively explain this
observation by the associated decrease in electronic charge density of the
intial $\sigma_g$-orbital after the polarization direction (see countour
plot in Fig.~\ref{fig:rotation}). The same reasoning carries over to the
behavior of the total alignment dependent ionization probabilities shown in
Fig.~\ref{fig:ion_tot}. The results in this figure can be obtained by
integrating the differential ionization probability
Eq.~\eqref{eq:diff_ionization} over all directions.  Alternatively, we may
project out the bound state components of the final wave function.  For
comparison, Fig.~\ref{fig:ion_tot} also contains the results from
Ref.~\cite{kamta:2005} which were obtained by a field of the same frequency
and peak intensity but with a slightly different pulse shape (trapezoidal)
and longer duration. We find somewhat lower ionization probabilities than in
Ref.~\cite{kamta:2005} since our pulse is at the peak intensity for a
shorter duration of time. Although the two data series are not directly
comparable, the overall behaviour is similar, namely decreasing ionization
probability with increasing alignment angle from parallel ($0^\circ$) to
perpendicular ($90^\circ$).

\section{Conclusion and outlook}
\label{sec:conclusion}

In conclusion, we have developed a new approach that accurately and
efficiently resolves the  $m$-mixing problem in large scale computations in
a spherical coordinate system. The method relies on an identification of
rotations in the intermediate propagation that brings the wave function into
a frame of reference in which $m$ is conserved. This means that
time-consuming $m$-mixing induced by the external perturbation is avoided
and instead delegated to the rotations which are very efficiently
implemented using the Wigner rotation matrix representation of the rotation
operator in the spherical harmonics basis.

We have chosen the linear molecule interacting with a linearly polarized
field to illustrate our method, but a similar approach can be used in a much
broader range of three-dimensional problems.  For example we could consider
an elliptically polarized field. In the split operator method we take the
time step $\tau$ to be small enough such that the field can be taken to be
constant both in magnitude and polarization direction within the small time
interval. We can therefore consider a time-dependent laboratory frame which
follows the instantaneous polarization direction. If we make the
transformation from the molecular frame to the new laboratory frame, we are
again able to treat the field as being linear and propagate as discussed
above. Our method can also be extended to arbitrary nuclear positions. For
any nuclear configuration, we can attach a coordinate system to each nucleus
with a $z$ axis from the origin to the nucleus. Then we decompose the
molecular potential to a sum of nuclear potentials, each of which can be
propagated with azimuthal symmetry in their own reference frame. Despite the
fact that we now need rotations between the coordinate systems belonging to
all of the nuclei, the total calculation is still in the same complexity
class with respect to scaling in $l_\text{max}$.

This work is supported by the Danish Research Agency (Grant. No.
2117-05-0081).

\bibliographystyle{apsrev}

\end{document}